\documentclass[a4paper]{article}
\usepackage{graphicx}
\usepackage[compress]{cite}
\usepackage{braket}
\usepackage{bm}
\usepackage{comment}
\usepackage{here}
\usepackage{a4wide}
\usepackage{authblk}
\usepackage{amsmath,amssymb}
\usepackage[usenames,dvipsnames]{color} 
\usepackage[normalem]{ulem} 
\usepackage{verbatim}

\begin{document}
\def\Journal#1#2#3#4{{#1} {\bf #2}, #3 (#4)}
\def\AHEP{Advances in High Energy Physics.} 	
\def\ARNPS{Annu. Rev. Nucl. Part. Sci.} 
\def\AandA{Astron. Astrophys.} 
\def\ANP{Ann. Phys.}
\def\APJ{Astrophys. J.}
\def\APJL{Astrophys. J. Lett.}
\def\APJS{Astrophys. J. Suppl}
\def\CMP{Commn. Math. Phys.}
\def\COMR{Comptes Rendues}
\def\CQG{Class. Quantum Grav.}
\def\CPC{Chin. Phys. C}
\def\EPJC{Eur. Phys. J. C}
\def\EPL{EPL}
\def\FP{Fortsch. Phys.}
\def\IJMPA{Int. J. Mod. Phys. A}
\def\IJMPE{Int. J. Mod. Phys. E}
\def\JCAP{J. Cosmol. Astropart. Phys.}
\def\JHEP{J. High Energy Phys.}
\def\JETPL{JETP. Lett.}
\def\JETPUSSR{JETP (USSR)}
\def\JPG{J. Phys. G} 
\def\JPCS{J. Phys. Conf. Ser.} 
\def\JPGNP{J. Phys. G: Nucl. Part. Phys.} 
\def\MNRAS{Mon. Not. R. Astron. Soc.}
\def\MPLA{Mod. Phys. Lett. A}
\def\NIMA{Nucl. Instrum. Meth. A.}
\def\NATU{Nature}
\def\NCA{Nuovo Cimento}
\def\NJP{New. J. Phys.}
\def\NPB{Nucl. Phys. B}
\def\NPBOLD{Nucl. Phys.}
\def\NPBSUPPL{Nucl. Phys. B. Proc. Suppl.}
\def\PDU{Phys. Dark. Univ.}
\def\PL{Phys. Lett.}
\def\PLB{{Phys. Lett.} B}
\def\PMCA{PMC Phys. A}
\def\PREP{Phys. Rep.}
\def\PPNP{Prog. Part. Nucl. Phys.}
\def\PLBOLD{Phys. Lett.}
\def\PAN{Phys. Atom. Nucl.}
\def\PRL{Phys. Rev. Lett.}
\def\PRD{Phys. Rev. D}
\def\PRC{Phys. Rev. C}
\def\PR{Phys. Rev.}
\def\PTP{Prog. Theor. Phys.}
\def\PTEP{Prog. Theor. Exp. Phys.}
\def\RMP{Rev. Mod. Phys.}
\def\RPP{Rep. Prog. Phys.}
\def\SJNP{Sov. J. Nucl. Phys.}
\def\SPJETP{Sov. Phys. JETP.}
\def\SCIENCE{Science}
\def\TNYAS{Trans. New York Acad. Sci.}
\def\ZETP{Zh. Eksp. Teor. Piz.}
\def\ZFPH{Z. fur Physik}
\def\ZPC{Z. Phys. C}
\title{Reply to ``Comment on `Unified neutrino mixing and approximate $\mu-\tau$ reflection symmetry'[arXiv:2603.00885]''}
\author{Yuta Hyodo \footnote{yuta.h.1410@gmail.com; Formerly with Tokai University}}
\author[2]{Teruyuki Kitabayashi \footnote{teruyuki@tokai.ac.jp}}
\affil[2]{Department of Physics, School of Science, Tokai University, 4-1-1 Kitakaname, Hiratsuka, Kanagawa 259-1292, Japan}
\date{}
\maketitle

\begin{abstract}
Huang and Li [arXiv:2603.00885] have raised the following two points regarding our previous work [arXiv:2502.18029]:(1) The real-value conditions associated with $\mu$-$\tau$ reflection symmetry were overlooked. (2) Inverted Ordering (IO) remains viable when the latest experimental data are taken into account. As they have pointed out, we overlooked an important real-value condition. However, with regard to point (2), we believe that there may be a misunderstanding. In our original study, we excluded IO based on the constraint on the sum of neutrino masses, $\sum m_\nu$. In contrast, they argue that IO remains viable when considering the effective neutrino mass, $|M_{ee}|$. While IO may indeed remain allowed in light of the latest $|M_{ee}|$ data, it is still in tension with the experimental bounds on $\sum m_\nu$ under approximate $\mu-\tau$ symmetry within the discussed model parameter space.

\end{abstract}

\vspace{1cm}

Huang and Li ~\cite{Huang:2026} have raised the following two points regarding our previous work~\cite{Hyodo:2025}:
\begin{description}
    \item[(1)] The real-value conditions associated with $\mu$--$\tau$ reflection symmetry were overlooked.
    \item[(2)] Inverted Ordering (IO) remains viable when the latest experimental data are taken into account.
\end{description}
We appreciate the efforts of Huang and Li in carefully examining our work and providing their observations.

First, we would like to express our sincere gratitude to Huang and Li for their insightful comment in point (1). As they have correctly pointed out, we indeed overlooked an important real-value condition. However, with regard to point (2), we believe that there may be a misunderstanding.

In our original study, we excluded IO based on the constraint on the sum of neutrino masses, $\sum m_\nu$. In contrast, they argue that IO remains viable when considering the effective neutrino mass, $|M_{ee}|$. While IO may indeed remain allowed in light of the latest $|M_{ee}|$ data, it is still in tension with the experimental bounds on $\sum m_\nu$.

They corrected our calculations by incorporating the real-value conditions---the results of which are displayed in Figure~1 of \cite{Huang:2026}---we suggest that the vertical lines representing the allowed range of $\sum m_\nu$ (as shown in Figures~1 and 2 of \cite{Hyodo:2025}) should also be overlaid on their Figure~1. With this addition, one can see that even after properly taking into account the previously overlooked real-value conditions, IO is still excluded.

Consequently, the main conclusion of our paper~\cite{Hyodo:2025}---that IO is excluded under approximate $\mu-\tau$ symmetry within the discussed model parameter space---has actually been further strengthened thanks to the refinements provided by Huang and Li.

\vspace{1cm}

\end{document}